\documentclass{aastex}
\usepackage{emulateapj5}
\usepackage{amssymb}
\makeatletter

\renewcommand{\cite}[1]{\citeyear{#1}}
\bibpunct{\hspace{-4pt}}{}{}{a}{}{}
\submitted{{\rm Accepted to ApJ Lett.}}
\newcounter{twocolfigure}[figure]
\newcommand{\twocolfigurecap}[2]{%
   \refstepcounter{twocolfigure}%
   \label{#1}%
   {\small{\sc%
   Fig.} \thetwocolfigure --- #2}%
   \vspace{1ex}%
}
\makeatother
\begin{document}

\newcommand{\Ua}{\widetilde{a}}

\newcommand{\UP}{\widetilde{P}}

\newcommand{\Um}{\widetilde{m}}

\newcommand{\Ut}{\widetilde{t}}

\newcommand{\Urp}{\widetilde{r}_{p}}

\newcommand{\UdE}{\Delta\widetilde{E}}

\newcommand{\Urt}{\widetilde{r}_{t}}

\newcommand{\Urh}{\widetilde{r}_{h}}

\newcommand{\Urc}{\widetilde{r}_{c}}

\newcommand{\UE}{\widetilde{E}}

\newcommand{\UR}{\widetilde{R}}

\newcommand{\Mo}{M_{\odot}}

\newcommand{\Ro}{R_{\odot}}

\newcommand{\Lo}{L_{\odot}}

\newcommand{\HLm}{\widehat{L}_{0}}

\newcommand{\Utm}{\widetilde{t}_{0}}

\newcommand{\UT}{\widehat{T}}

\newcommand{\ULt}{\widetilde{L}_{t}}

\newcommand{\HLt}{\widehat{L}_{t}}

\newcommand{\Utd}{\widetilde{\tau}_{d}}

\newcommand{\Ms}{M_{\star}}

\newcommand{\Rs}{R_{\star}}

\newcommand{\Ls}{L_{\star}}

\newcommand{\Ts}{T_{\star}}

\newcommand{\Ks}{K_{\star}}

\newcommand{\Ns}{N_{\star}}

\newcommand{\UWp}{\widetilde{\Omega}_{p}}

\newcommand{\UdJ}{\Delta\widetilde{J}}

\newcommand{\Te}{T_{\mathrm{eff}}}

\title{Squeezars: Tidally powered stars orbiting a massive black hole}

\author{Tal Alexander}

\affil{Faculty of Physics, The Weizmann Institute of Science, POB 26, Rehovot
76100, Israel}

\author{\and}

\author{Mark Morris}

\affil{Division of Astronomy and Astrophysics, University of California,
Los Angeles, CA 90095-1562, USA}

\begin{abstract}
We propose that there exists a class of transient sources, {}``squeezars'',
which are stars caught in highly eccentric orbits around a massive
($m\!\lesssim\!10^{8}\,\Mo$) black hole (MBH), whose atypically high
luminosity (up to a significant fraction of their Eddington luminosity)
is powered by tidal interactions with the MBH. Their existence follows
from the presence of a mass sink, the MBH, in the galactic center,
which drives a flow of stars into nearly radial orbits to replace
those it has destroyed. We consider two limits for the stellar response
to tidal heating: surface heating with radiative cooling ({}``hot
squeezars'') and bulk heating with adiabatic expansion ({}``cold
squeezars''), and calculate the evolution of the squeezar orbit,
size, luminosity and effective temperature. The squeezar formation
rate is only $\sim\!0.05$ that of tidal disruption flares, but squeezar
lifetimes are many orders of magnitude longer, and so future observations
of squeezars in nearby galaxies can probe the tidal process that feeds
MBHs and the effects of extreme tides on stars. The mean number of
squeezars orbiting the Galactic MBH is estimated at $0.1$--$1$.
\end{abstract}


\section{Introduction}

A massive galactic black hole is a mass sink that drives an inflow
of stars from its radius of influence, $r_{h}$. When the MBH mass
$m$ is small enough ($m\!\lesssim\!10^{8}\,\Mo$ for solar type stars),
the tidal disruption radius $r_{t}$ lies outside the event horizon
and the star is disrupted before falling in. Prompt disruption occurs
when stars are scattered into highly eccentric (loss-cone) orbits
with periapse $r_{p}\!<\! r_{t}$. The accretion of the debris may
be observed as a short-lived ($\sim\!1\,\mathrm{yr}$) luminous {}``tidal
flare'' (Frank \& Rees \cite{Fra76}). 

Stars with $r_{p}\!\gtrsim\! r_{t}$ narrowly escape after undergoing
extreme tidal distortion, mass-loss, spin-up and subsequent mixing,
which may affect their evolution and appearance. The near-miss event
rate is comparable to that of tidal disruptions (Alexander \& Livio
\cite{Ale01b}). Most of these stars avoid tidal capture by being
deflected to wider orbits, or by missing the MBH due to its Brownian
motion. Such {}``tidally scattered'' stars eventually amount to
a few percent of the stellar population within $r_{h}$. The rest
are tidally captured and gradually spiral in as the tides convert
a small fraction of their orbital energy into heat each peri-passage
(Alexander \& Hopman \cite{AH03}, AH03). 

The orbital energy a star must lose to circularize far exceeds its
own binding energy. A {}``squeezar'', a tidally heated star orbiting
a MBH, is ultimately disrupted by expanding beyond its Roche lobe
or by radiating above its Eddington luminosity (Rees \cite{Ree88};
Novikov, Petchik \& Polnarev \cite{Nov92}). Squeezars directly trace
the tidal disruption process, which is important for feeding low-mass
MBHs. The squeezar phase lasts orders of magnitude longer than a tidal
flare, and so observations of squeezars in nearby galaxies, in particular
IR observations of the Galactic Center (GC), can probe the tidal disruption
process and the effects of extreme tides on stars.

\section{Squeezar evolution }

\label{sec:evol}

The evolution of a squeezar reflects its structure and the way its
mechanical and thermal properties respond to tidal heating. We approach
the challenging problem of modeling the evolution by considering two
simple scenarios that likely bracket the range of possible responses:
(1) Surface heating and radiative cooling ({}``hot squeezar'', HS),
where the tidal oscillations dissipate in a very thin surface layer
that expands moderately and radiates at a significantly increased
effective temperature, $\Ts$ (McMillan, McDermott \& Taam \cite{McM87}).
(2) Bulk heating and adiabatic expansion ({}``cold squeezar'', CS),
where the oscillations dissipate in the stellar bulk and cause a large
quasi adiabatic, self-similar expansion at constant $\Ts$ (Podsiadlowski
\cite{Pod96}). 

We consider the orbital decay of a star of mass $\Ms\!\ll\! m$ and
initial radius $\Rs$ that is tidally captured by a MBH in a region
where the MBH dominates the potential (Keplerian approximation). We
denote by a tilde quantities in dimensionless units of $G$=$\Ms$=$\Rs$=1.
Thus time is measured in terms of the stellar dynamical time $t_{\star}\!\equiv\!\sqrt{\Rs^{3}/G\Ms}\!\simeq\!1600\,\mathrm{s}\,(\Rs/\Ro)^{3/2}(\Ms/\Mo)^{-1/2}$
and energy in terms of the stellar binding energy (up to a factor),
$E_{\star}\!\equiv\! G\Ms^{2}/\Rs\!\simeq\!3.8\!\times\!10^{48}\,\mathrm{erg}\,(\Ms/\Mo)^{2}(\Rs/\Ro)^{-1}$. 

The tidal disruption radius, $\Urt\!\simeq\!\UR(\Ut)\Um^{1/3}$ ($\UR$
is the stellar radius), lies outside the event horizon of a non-rotating
MBH when $\Um\!<\!(\widetilde{c}/\sqrt{2})^{3}$($\widetilde{c}$
is the speed of light). The semi-major axis, period and eccentricity
of a Keplerian orbit are related to the orbital energy $\UE$ by\begin{equation}
\Ua=-\Um/2\UE\,,\:\UP=2\pi\sqrt{\Ua^{3}/\left(1+\Um\right)}\,,\: e=1-\Urp/\Ua\,.\label{eq:orb}\end{equation}

If orbital angular momentum is conserved, the circularization radius
is $\Urc\!=\!2\Urp$. Conversely, if the angular momentum transfered
from the orbit to the tides follows the impulsive relation $\UdJ\!=\!\UdE/\UWp$,
where $\UWp$ is the orbital angular velocity at periapse (e.g. Kumar
\& Quataert \cite{Kum98}), then $\Urp\!=\!\Urc\!=\!\mathrm{const}$.
In either case $\Urp\!\sim\!\mathrm{const}$ in the early stages of
circularization. We assume here $\Urp\!=\!\mathrm{const}$, so the
orbital energy the star has to lose in order to circularize from an
$\UE\!\sim\!0$ orbit is $\UE_{c}=\Um\left/2\Urp\right.=\Um^{2/3}\left/2b\right.\gg1$. 

The tidal energy extracted from the orbit in a single peri-passage
is given to leading order in the linear multipole expansion by (e.g.
Press \& Teukolsky \cite{Pre77}) \begin{equation}
\UdE_{t}=T_{2}\left(\eta\right)\UR^{5}/b^{6}\,,\label{eq:dEt}\end{equation}
where $b\!\equiv\!\Urp/\Urt(\Ut\!=\!0)$ and the $\UR^{5}$ term accounts
for possible stellar expansion. The 2nd order tidal coupling coefficient
$T_{2}$ depends on the stellar structure, on $e$, and on the dimensionless
transit time, $\eta\equiv\Urp^{3/2}/\sqrt{1+\Um}\simeq b^{3/2}$ ($\Um\!\gg\!1$).
The linear mode analysis is formally valid for $\eta\!>\!1$, which
holds here. We assume here that $e\!=\!1$ since the $P$--$e$ relation,
$1\!-\! e\simeq(2\pi/\UP)^{2/3}b$, indicates that $e\!\simeq\!1$
down to $P\!\sim\!0.1\,\mathrm{yr}$ (e.g. at that point $1\!-\! e\!=\!0.02b$
for a solar type star), and most squeezars are disrupted well before
reaching such short periods (\S\ref{sec:GCsqueezars}).

The tidal coupling changes as the stellar structure and spin evolve.
The analysis below is greatly simplified by assuming that $T_{2}$
is constant in time. This is suggested by the fact that the squeezar
phase coincides with the initial stages of circularization and synchronization.
In the case of bulk heating (\S\ref{sub:ColdSq}), the star expands
roughly self-similarly and so the structural changes can be accounted
for by the $\UR^{5}$ term, while in the case of surface heating the
expanding layer involves only a minute fraction of the stellar mass
and the stellar bulk is unaffected (\S\ref{sub:HotSq}). We do not
consider possible resonances between the orbital period and the tidal
oscillations (Novikov et al. \cite{Nov92}). This is not relevant
for CSs, where the tidal energy is carried by high order modes and
quickly dissipated. For HSs, we note that small orbital perturbations
($\delta P$, $\delta E$) by other stars will randomize the phase
between the orbit and the tidal oscillations as long as $\tau/P\!<$$\delta P/P$=$3\delta E/2E\!\sim$$3P/2t_{r}$,
where $\tau\!<\! t_{\star}$ is the width of the resonance, and $t_{r}$
is the 2-body relaxation time. E.g., for $\tau\!=\!0.1t_{\star}\!\sim\!100\,\mathrm{s}$
and $t_{r}\!=\!10^{9}\,\mathrm{yr}$, resonances are suppressed for
$P\!\gtrsim\!50$ yr. Further analysis is needed to validate these
assumptions.

Once the details of the tidal energy deposition and the stellar response
are specified, Eqs. (\ref{eq:orb}) and (\ref{eq:dEt}) can be evolved
numerically orbit by orbit. We now derive approximate analytic solutions
for the evolution of squeezars.

\subsection{Hot squeezars: surface heating and radiative cooling}

\label{sub:HotSq}

The place where the tidal oscillations thermalize depends on the modes
that carry the energy. In stars with large convective envelopes, it
is mainly carried by $f$ and $p$ modes that, absent non-linear couplings
to higher modes, dissipate in the outermost layers (e.g. McMillan
et al. \cite{McM87}). We adopt this as a limiting case, and assume
that $\UR\!=\!1$ in Eq. (\ref{eq:dEt}) since the expansion involves
only a thin surface layer with a minute fraction of the stellar mass.
The implied assumptions are that the oscillations occur in the unchanged
stellar bulk just below the surface layer in which they dissipate,
and that the oscillatory modes are unaffected by the modified surface
boundary conditions.

The orbital evolution of a HS is then derived from the relation $d\UP/\UP\!\simeq\!(d\UP/d\UE)\UdE_{t}/\UP$
(assuming $|\UdE_{t}/\UE|\!\ll\!1$, as is the case here), whose solution
is \begin{equation}
\UP=\UP_{0}(1-\Ut/\Utm)^{3}\,,\qquad\Ua=\Ua_{0}(1-\Ut/\Utm)^{2}\,,\label{eq:Pt}\end{equation}
where $\UP_{0}$ and $\Ua_{0}$ are the initial values and where \begin{equation}
\Utm=[4\pi^{2}\UP_{0}/(1+\Um)]^{1/3}\Um/\UdE_{t}=2\UE_{0}/(\mathrm{d}\UE/\mathrm{d}\Ut)_{0}\end{equation}
 The formal result $\UP(\Utm)\!=\!0$ is an artifact of the assumption
$\UdE_{t}\!\sim\!\mathrm{const}.$ If the star could avoid expansion
and tidal disruption, then ultimately $\UdE\!\rightarrow\!0$ (Hut
\cite{Hut80}) and $\UP\!\rightarrow\!2\pi b^{3/2}$. Because the
final stages of the stellar expansion are rapid, $\Utm$ also estimates
the total time to disruption for $b\!\gtrsim\!1.5$. The number of
orbits completed by time $\Ut$ is $N_{\mathrm{orb}}=\int_{0}^{\Ut}d\Ut^{\prime}/\UP(\Ut^{\prime})=(\Utm/2\UP_{0})\left[(1-\Ut/\Utm)^{-2}-1\right]$.

The energy in the stellar oscillations is dissipated and radiated
over a timescale $\Utd$, which can span several orders of magnitude
depending on the stellar structure and the nature of the excited mode,
but usually $\Utd\!>\!\UP$. Here we adopt a typical value of $\tau_{d}\!\sim\!10^{4}\,\mathrm{yr}$
(Ray, Kembhavi \& Antia \cite{Ray87}; McMillan et al. \cite{McM87}).
We assume that the tidal luminosity $\ULt$ (the luminosity in excess
of the initial luminosity $\Ls$) that is released by one peri-passage
decays exponentially, $\ULt=(\UdE_{t}/\Utd)\exp(-\Ut/\Utd)$, where
$\Ut$ is the time after periapse. The luminosity from many successive
peri-passages is given by the relation \mbox{$\ULt(\Ut+\UP)=[\ULt(\Ut)+\UdE_{t}/\Utd]\exp(-\UP/\Utd)$}.
This can be solved approximately by substituting $\UP\!=\!\mathrm{const}$,
since most of the luminosity comes from the recent peri-passages that
have roughly the same period. The luminosity is then\begin{equation}
\ULt=(\UdE_{t}/\Utd)\left[1\!-\!\exp(-\Ut/\Utd)\right]\left/\left[\exp(\UP/\Utd)\!-\!1\right]\right.\,,\label{eq:Lt}\end{equation}
for the initial condition $\ULt(0)\!=\!0$. At later times, when $\Ut\gg\Utd$
and $\UP\ll\Utd$, $\ULt\!\rightarrow\!\UdE_{t}/\UP$ irrespective
of $\Utd$.

Numeric models of tidal heat dissipation in a thin surface layer (McMillan
et al. \cite{McM87}; Podsiadlowski \cite{Pod96}) indicate that the
stellar interior is essentially unaffected and that in the limit $\HLt\!\gg\!1$
($\HLt\!\equiv\! L_{t}/\Ls$), the binding energy required to raise
the expanded layer above the stellar surface can be empirically expressed
as a power-law of the tidal luminosity, $\UdE_{\star}\propto1-1/\UR\propto\HLt^{\alpha}$.
Therefore \begin{equation}
\UR^{-1}=1-(\HLt/\HLm)^{\alpha}\qquad(\HLt\gg1)\,,\label{eq:Rt}\end{equation}
 where $\HLm$ is the terminal luminosity for which $\UR\!\rightarrow\!\infty$.
The results of McMillan et al. (\cite{McM87}) indicate that $\alpha\!\sim\!0.2$
for a $0.8\, M_{\odot}$ main sequence (MS) star, $\alpha\!\sim\!0.4$
for a $1.5\, M_{\odot}$ star, and that $2\!\times\!10^{3}\!<\!\HLm\!<\!10^{4}$,
depending on the depth of the heated layer (smaller values correspond
to deeper layers). The terminal luminosity $\HLm$ is of the order
of the stellar Eddington limit $\widehat{L}_{\mathrm{E}}\!=\!3.2\!\times\!10^{4}(\Ms/\Mo)(\Ls/\Lo)^{-1}$.
The stellar expansion is truncated when $\UR\!=\! b$, and the tidal
disruption radius overtakes the star. From that point on, the stellar
lifetime is limited by mass loss at periapse. Here we conservatively
assume prompt destruction when $\UR\!=\! b$. 

The change in the effective temperature, $\UT\!\equiv\! T/\Ts$, is
\begin{equation}
\UT^{4}=(1+\HLt)/\UR^{2}\,.\label{eq:Tt}\end{equation}
Since $\HLt\!\le\!\HLm$ while $\UR$ diverges, there exists a maximal
temperature, which is attained when $\UR\!=\!1\!+\!1/2\alpha$, or
equivalently when $\HLt/\HLm\!=\!(1+2\alpha)^{-1/\alpha}$ ($\HLt\!\gg\!1$),
\begin{equation}
\max\UT\simeq\sqrt{2\alpha}(1+2\alpha)^{-(1+2\alpha)/4\alpha}\HLm^{1/4}\sim\!0.4\HLm^{1/4}\,.\end{equation}
For the $\alpha$ and $\HLm$ range considered here, $2.3\!<\!\max\UT\!<\!4.6$.
For $\Ts\!\gtrsim\!10^{4}\,\mathrm{K}$, the change in the stellar
K-band magnitude can be roughly estimated by the Rayleigh-Jeans limit,
\begin{equation}
\Delta K=K-\Ks\sim-2.5\log_{10}(\UR^{2}\UT)\,,\label{eq:Kt}\end{equation}
where $\Ks$ is the star's initial magnitude. The actual magnitude
is $\sim\!0.5^{\mathrm{m}}$ brighter for a solar-type star (by numeric
integration of the blackbody spectrum; the values quoted below are
exact). The $K$-luminosity increases with the stellar radius and
so is largest just before disruption. At maximal temperature, $-4.1^{\mathrm{m}}\!\lesssim\!\Delta K\!\lesssim\!-3.0^{\mathrm{m}}$. 

These analytic expressions fully describe the evolution of a HS given
initial orbital parameters ($b$ and $\UP_{0}$) and stellar type
($\Ms$, $\Rs$, $\Ls$, $\Ts$ and $\Ks$). $T_{2}$ can be evaluated
numerically for a given stellar model. Here we represent a typical
star by the Sun, and calculate $\UdE_{t}$ for a detailed solar model
(Alexander \& Kumar \cite{Ale01a}). Figure \ref{f:evol} confirms
that the exact orbit-by-orbit evolution of HSs (\S\ref{sec:evol})
is well approximated by Eqs. (\ref{eq:Pt}--\ref{eq:Kt}).

\subsection{Cold squeezars: bulk heating and adiabatic expansion}

\label{sub:ColdSq}

Stars subjected to high amplitude tidal deformations may dissipate
the tidal energy in their bulk via non-linear mode couplings, which
redistribute the energy among an infinitude of high order modes (Kumar
\& Goodman \cite{Kum96}). These dissipate very quickly, up to $10^{6}$
times faster than low order modes, and so bulk heating is effectively
instantaneous. Numeric modeling of bulk heated stars (Podsiadlowski
\cite{Pod96}) indicates that they expand quasi adiabatically and
self similarly on a dynamical timescale while maintaining their original
effective temperature. 

The tidal heat is stored in the stellar binding energy,\begin{equation}
\UdE_{\star}=\beta(1-1/\UR)\,,\end{equation}
where the dimensionless factor $\beta$ depends on the stellar mass
concentration and heat capacity ratio. Typically $\beta\!=\!0.75$
for a MS star (e.g. Cox \& Giuli \cite{Cox68}). The tidal heat is
radiated from the expanded surface, and so the luminosity peaks just
before tidal disruption, when $\widehat{L}\!=\! b^{2}$. Since typically
$\HLt\!<\!10$, CSs are far less luminous than hot ones, but because
$\Delta K$ is dominated by the stellar expansion, they brighten by
up to $-5\log b\!\sim\!-2.5^{\mathrm{m}}$, only $1^{\mathrm{m}}$
less than HSs. We adopt the bulk heating scenario as a limiting case,
and estimate CS evolution with the simplifying assumption of adiabatic
expansion, where the radiated luminosity is neglected in the energy
budget. The orbital energy, $\UE\!=\!\UE_{0}\!-\!\UdE_{\star}$, then
depends only on $\UR$ and on the initial orbital energy $\UE_{0}$.
The adiabatic assumption fails for some combinations of orbital parameters,
when $\ULt=\widetilde{L}_{\star}(\UR^{2}-1)\!\gtrsim\!\UdE_{t}/\UP$.
An adiabatic CS is disrupted earlier than a radiative one because
it expands more efficiently, and because the $\UR^{5}$ dependence
of $\UdE_{t}$ (Eq. \ref{eq:dEt}) accelerates the heating rate. A
radiative CS evolves more slowly, and so takes longer to reach its
high luminosity phase, but then spends more time there. Therefore,
adiabatic bulk heating provides a lower limit on CS lifetimes. 

The adiabatic evolution of a CS, derived from $d\UdE_{\star}/d\Ut\!=\!\beta(d\UR/d\Ut)/\UR^{2}\!=\!\UdE_{t}/\UP(\UR)$,
is \begin{equation}
\widetilde{t}(\widetilde{R}_{2})-\widetilde{t}(\widetilde{R}_{1})=\Theta(b,\varepsilon_{0})[f(x_{1})-f(x_{2})]\,,\label{eq:dt}\end{equation}
where $\varepsilon_{0}\equiv-\widetilde{E}_{0}/\beta$, $x\equiv(1+\varepsilon_{0})\widetilde{R}$,
and where 

\begin{equation}
\Theta(b,\varepsilon_{0})\equiv\frac{256\sqrt{2}\pi\Um^{3/2}b^{6}(1+\varepsilon_{0})^{9/2}}{63\sqrt{\beta(1+\Um)}T_{2}(b^{3/2})}\,,\end{equation}
 \begin{equation}
f(x)\equiv\frac{-7\!-\!10x\!-\!16x^{2}\!-\!32x^{3}\!-\!128x^{4}\!+\!256x^{5}}{256\sqrt{x^{9}(x-1)}}\,.\end{equation}
Neglecting tidal disruption, the star expands to infinity in \begin{equation}
\Utm=\Theta(b,\varepsilon_{0})[f(1+\varepsilon_{0})-1]\,.\end{equation}
 Since the latter stages of the expansion are rapid, $\Utm$ estimates
the total time to disruption for $b\gtrsim\!1.5$. Conservative upper
limits on the terminal values of the orbital parameters at disruption
are ($\varepsilon_{0}\ll1$), 

\begin{equation}
\begin{array}{ccc}
\widetilde{a} & \simeq & \widetilde{m}b/[2\beta\left(b-1\right)]\\
e & \simeq & 1-2\beta\left(b-1\right)/\widetilde{m}^{2/3}\\
\widetilde{P} & \simeq & \pi\left\{ \widetilde{m}b/[\beta(b-1)]\right\} ^{3/2}/\sqrt{2(1+\Um)}\end{array}\,.\label{eq:aePt}\end{equation}

\section{Squeezars in the Galactic Center}

\label{sec:GCsqueezars}

At a distance of $8\,\mathrm{kpc}$ (Reid \cite{Rei93}), the $\sim\!3\!\times\!10^{6}\,\Mo$
MBH in the GC (Ghez et al. \cite{Ghe00}; Sch\"odel et al. \cite{Sch02})
is the nearest and most accessible MBH. Although it is heavily reddened
($A_{K}\!\sim\!3^{\mathrm{m}}$, Blum et al. \cite{Blu96}), deep
high resolution IR observations provide information on the luminosity,
temperature and orbits of thousands of stars near the MBH (Eckart
et al. \cite{Eck99}; Figer et al. \cite{Fig00}; Gezari et al. \cite{Gez02}).
The squeezar formation rate in the GC is $\Gamma\!\sim\!5\!\times10^{-6}\,\mathrm{yr}^{-1}$,
only $\sim\!0.05$ of the prompt tidal disruption rate (AH03). However,
squeezars are relatively long-lived, and so on average $\overline{n}\!\sim\!0.1$--$1$
squeezars orbit the MBH at any given time. The leading (shortest period)
squeezar has typically completed $\Ut/\overline{t}_{0}\sim\!\overline{n}/(\overline{n}+1)$
of its lifetime, where $\overline{t}_{0}$ is the mean inspiral time.

The typical properties of the leading squeezar can be derived by averaging
over all initial orbits, weighted by the probability of successful
inspiral, which falls with increasing periapse and initial orbital
period (AH03). For $1\,\Mo$ hot squeezars in the GC, this yields
$L_{t}\!\sim\!170\,\Lo$, $T_{\mathrm{eff}}\!\sim\!19000\,\mathrm{K}$,
$\Delta K\!\sim\!-2.25$, $P\!\sim\!3.6\!\times\!10^{3}\,\mathrm{yr}$
and $1\!-\! e\!\sim\!2\!\times\!10^{-5}$. Figure \ref{f:evol} shows
the evolution of a HS up to disruption. The CS orbital bounds (Eq.
\ref{eq:aePt}) are similar. Note that the luminosity and temperature
may be much higher if $\overline{n}$ was even modestly under-estimated.

\section{Discussion}

The existence of squeezars near a MBH is a consequence of the flow
of stars into a mass sink. We approximated the complex evolution of
tidally heated stars by two simple models that likely bracket the
response of real stars and can serve as reference for future work. 

This study of squeezars is motivated by their potential to probe observationally
two important physical processes: MBH growth by tidal disruption and
the effects of strong tides on stars. Their advantage is that they
last orders of magnitude longer than either a tidal flare or a tidal
capture event in stellar binaries (e.g. McMillan et al. \cite{McM87}).

It is unclear whether squeezars are spectrally distinct from normal
stars, but their orbits should stand out since they require an extremely
small periapse. The colors of stars on highly eccentric orbits can
differentiate between surface and bulk heating, which are both plausible.
Two early type stars on orbits with $r_{p}\!\sim\!120$ AU, $e\!=\!0.87$
and $P\!=\!15.2$ yr (Sch\"odel et al. \cite{Sch02}) and $r_{p}\!=\!60$
AU, $e\!=\!0.98$ and $P\!=\!60$ yr (Ghez et al. \cite{Ghe03}) were
already discovered in the GC. Although $b\!\sim\!35$-$70$ is still
$>\!10$ too large for a squeezar, their detection shows that squeezar
searches are feasible with deep ($K\!\sim\!18\mathrm{^{m}}$, cf Fig.
\ref{f:evol}) high resolution adaptive optics (Genzel et al. \cite{Gen03}).
A few stars on tight eccentric orbits around the MBH are to be expected.
For a simple GC model with isotropic velocities (AH03), we estimate
that there should be $\sim\!120$ stars on $r_{p}\!<\!120$ AU, $P\!<\!60$
yr orbits. We note that $\overline{n}$ could be substantially under-estimated,
since several possible scattering mechanisms were not included in
the modeling of AH03. This possibility further motivates the search
for squeezars in the GC. If $\overline{n}\!\gg\!1$, a cluster of
squeezars in a nearby galaxy may also be observable.

\acknowledgements{We thank C. Hopman for his help. TA is supported by ISF grant 295/02-1,
Minerva grant 8484 and a New Faculty grant by Sir H. Djangoly, CBE,
of London, UK. MM is supported by NSF grant 99-88397.}

\begin{minipage}[t]{1.00\columnwidth}\vspace{1ex}\includegraphics[%
  clip,
  scale=0.68]{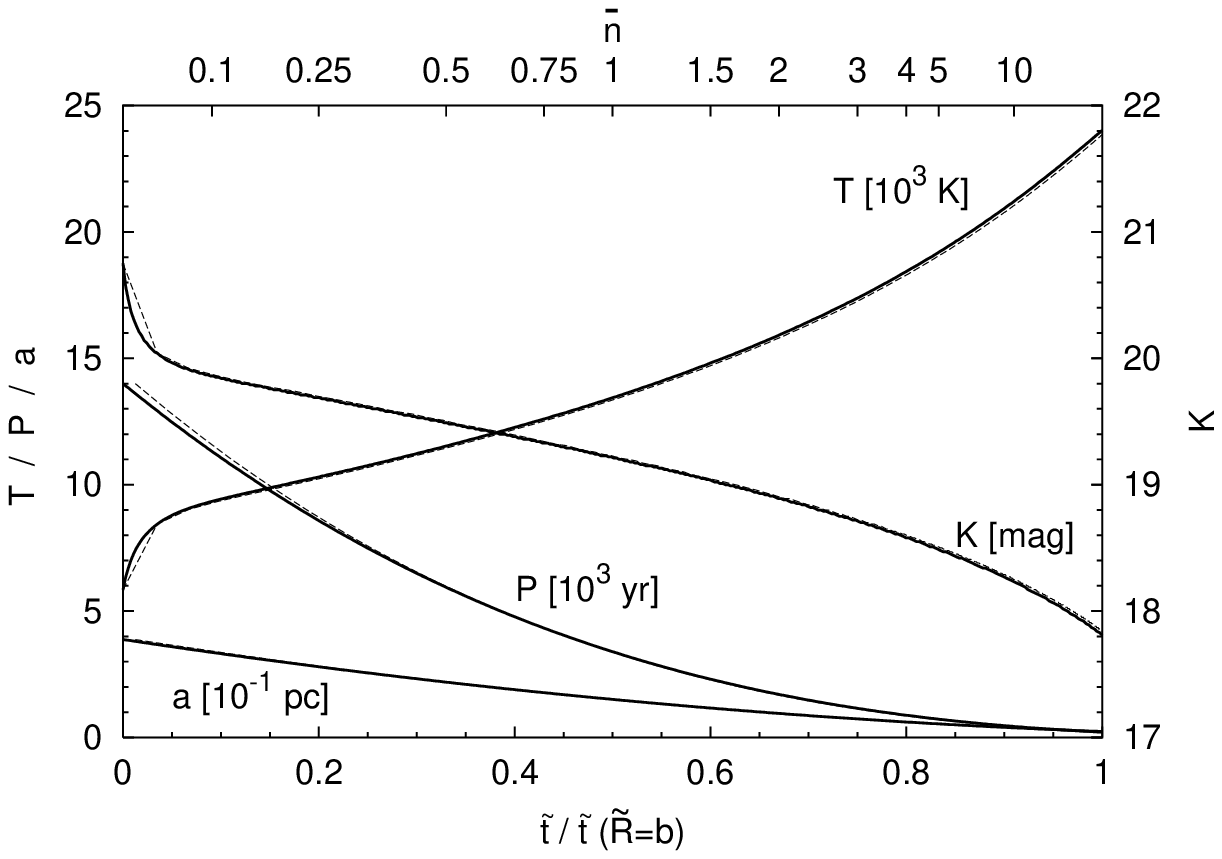}\vspace{1ex}\\ \twocolfigurecap{f:evol}{ The evolution of a $1\, M_{\odot}$
hot squeezar ($\alpha\!=\!0.4$, $\HLm\!=\!10^{4}$) in the GC, initially
deflected into a $b\!=\!1.5$, $P_{0}\!=\!1.4\!\times\!10^{4}$ yr
orbit ($t_{0}\!=\!4.9\!\times\!10^{5}$ yr), as calculated analytically
(full lines) and orbit by orbit (dashed lines). At disruption $t\!=\!3.7\!\times\!10^{5}$
yr, $L_{t}\!=\!642\,\Lo$, $P\!=\!210\,\mathrm{yr}$, $1\!-\! e\!=\!2.3\!\times\!10^{-4}$.
The leading squeezar properties are given in terms of $\overline{n}$,
assuming $\Ut_{0}\!\sim\!\overline{t}_{0}$ (top axis). }

\end{minipage}

\end{document}